\documentclass[aps,prb,twocolumn,superscriptaddress,showpacs]{revtex4}
\usepackage{graphicx,epsfig}
\usepackage{dcolumn}
\usepackage{bm}
\usepackage{amsmath,ulem,color}
\def\be{\begin{equation}}
\def\ee{\end{equation}}

\def\ber{\begin{eqnarray}}
\def\eer{\end{eqnarray}}
\def\kv{{\bm k}}

\def\qv{{\bm q}}

\begin{document}

\title{Persistent spin oscillations in a spin-orbit-coupled superconductor}
\author{Amit Agarwal}
\email{amit.agarwal@sns.it}
\affiliation{NEST, Scuola Normale Superiore and Istituto Nanoscienze-CNR, I-56126 Pisa, Italy}
\author{Marco Polini}
\affiliation{NEST, Istituto Nanoscienze-CNR and Scuola Normale Superiore, I-56126 Pisa, Italy}
\author{Rosario Fazio}
\affiliation{NEST, Scuola Normale Superiore and Istituto Nanoscienze-CNR, I-56126 Pisa, Italy}
\author{G. Vignale}
\affiliation{Department of Physics and Astronomy, University of Missouri, Columbia, Missouri 65211, USA}
\begin{abstract}
Quasi-two-dimensional superconductors with tunable spin-orbit coupling are very interesting systems with properties 
that are also potentially useful for applications. In this Letter we demonstrate that these systems exhibit undamped 
collective spin oscillations that can be excited by the application of a supercurrent. We propose to use these collective excitations to 
realize persistent spin oscillators operating in the frequency range of $10~{\rm GHz} - 1~{\rm THz}$.
\end{abstract}

\maketitle

{\it Introduction. ---} Spin-orbit-coupled two-dimensional (2D) electron gases (EGs) are the focus of great interest in the field of semiconductor spintronics~\cite{reviews}. This interest has been largely fueled by the hope to realize the visionary Datta-Das ``spin transistor"~\cite{datta_apl_1990} in which the on/off state is achieved by purely-electrical control of the electron's 
spin in a spin-orbit-coupled semiconductor channel placed between ferromagnetic leads. Research in spin-orbit-coupled 2DEGs has been recently revitalized by theoretical~\cite{SHE_theory} and experimental~\cite{SHE_experimental} studies of the spin Hall effect, in which a current traversing the sample generates a spin-current in the orthogonal direction.

The study of the interplay between spin-orbit coupling (SOC) and superconductivity in 2D systems,  stemming from  the seminal works of Edelstein~\cite{edelstein_prl_1995} and Gor'kov and Rashba~\cite{gorkov_prl_2001}, has also gained impetus~\cite{othergeneralrefs}. There is a large variety of systems in which SOC and superconductivity coexist:  two examples of great current interest are i) 2DEGs in InAs or GaAs semiconductor heterostructures that are proximized by ordinary s-wave superconducting leads~\cite{proximity2DEG,alicea_prb_2010} -- a class of systems which plays a key role in the quest for Majorana fermions~\cite{wilczek_natphys_2009} -- and ii) 2DEGs that form at interfaces between complex oxides~\cite{ohtomo_nature_2004}, such as ${\rm LaAlO}_3$ and ${\rm SrTiO}_3$, which display tunable SOC~\cite{caviglia_prl_2010} and superconductivity~\cite{caviglia_nature_2008}. 

Motivated by this body of experimental and theoretical literature, we investigate the collective spin dynamics of an archetypical 2DEG model Hamiltonian with Rashba SOC and s-wave pairing~\cite{gorkov_prl_2001}, in the presence of {\it repulsive} electron-electron (e-e) interactions. In the absence of superconductivity a Rashba 2DEG exhibits spin oscillations, which, at long wavelength and for weak repulsive interactions, have a frequency $\approx 2 \alpha k_{\rm F}$, $\alpha$ being the strength of SOC and $k_{\rm F}$ the 2D Fermi wavenumber in the absence of SOC. These oscillations, however, are damped and quickly decay due to the emission of (double) electron-hole pairs, which, in the normal phase, are present at arbitrary low energies.   In this Letter we demonstrate that in a Gor'kov-Rashba superconductor (GRSC), collective spin oscillations continue to exist  in a wide range of parameters, and  are undamped because they lie inside the superconducting gap where no other excitation exists. 
\begin{figure}
\centering
\includegraphics[width=0.9\linewidth]{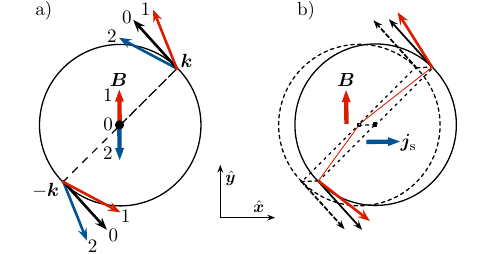}
\caption{(color online) a) Response of a Cooper pair in the $\lambda = +$ chirality subband of a Gor'kov-Rashba superconductor subjected to an oscillating  magnetic field in the ${\hat {\bm y}}$ direction.   The solid circle is the Fermi surface and the black dot is the origin of momentum space.  The arrows labeled by ``$1$",  ``$2$", and ``$0$" describe the orientation of the spins under the action of a magnetic field that points up, down, or vanishes.  Spontaneous oscillations  are sustained, in the absence of a magnetic field,  by the internal exchange field.  b)  A supercurrent  boosts the Fermi surface in the ${\hat {\bm x}}$ direction (solid line) and creates a magnetic field in the ${\hat {\bm y}}$ direction.  As a result, spins begin to oscillate around the new equilibrium orientation, indicated by the thick red arrows. \label{fig:one}}
\end{figure}
Fig.~\ref{fig:one} shows schematically the nature of the spin oscillations in a GRSC.  At variance with the Cooper pairs of a standard s-wave semiconductor,  the pairs of  a GRSC are in a mixture of singlet and triplet states.   It is this feature that enables the pairs to respond  to an oscillating magnetic field applied, say, in the ${\hat {\bm y}}$ direction.  In the course of the oscillation the spins of a  pair tilt in opposite directions, in a pair-breaking motion that creates a net spin polarization along the ${\hat {\bm y}}$ axis.  The spin polarization produces an exchange field, which, if the electron-electron interaction is sufficiently strong, sustains oscillations of the appropriate frequency in the absence of an external field. The essential point is that these oscillations are undamped as long as their frequency falls below the quasiparticle gap: they will therefore display an extraordinarily long lifetime~\cite{qfactor}.
 
In order to excite these long-lived spin modes one could in principle apply a short magnetic pulse, but there is also a purely-electrical method.  Namely, a supercurrent pulse applied, say, in the ${\hat {\bm x}}$ direction, will generate, {\it via} the Edelstein effect~\cite{edelstein_prl_1995} an effective magnetic field pulse in the ${\hat {\bm y}}$ direction, and this should be sufficient to start the spin oscillations.  This excitation mechanism is illustrated in Fig.~\ref{fig:one}b). The Fourier spectrum of the supercurrent pulse must not contain frequencies of the order of (or larger than) twice the superconducting gap to avoid the creation of quasiparticle excitations. We suggest that the new collective spin mode can be used to  realize ``persistent spin oscillators" operating in the frequency range of $10~{\rm GHz} - 1~{\rm THz}$ (for superconductors with a critical temperature in the range $10^{-1} - 10~{\rm K}$).

{\it Model Hamiltonian and effective low-energy theory.  ---} We consider the following model Hamiltonian: $\hat{\cal H} = \hat{\cal H}_{0} + \hat{\cal H}_{\rm p} + \hat{\cal H} _{\rm e-e}$.  Here $\hat{\cal H}_{0}$ is the kinetic energy term given by $\hat{\cal H}_{0} = \sum_{i, j} \int d^2{\bm r}~{\hat\psi}^{\dagger}_{i} ({\bm r})~h_{ij}({\bm r})~\hat{\psi}_{j} ({\bm r})$, where ($\hbar =1$ throughout this manuscript)
\be \label{eq:H0}
h_{ij}({\bm r}) = \frac{(-i {\bm \nabla}_{\bm r})^{2}}{2 m} \delta_{{ij}} + \alpha~[{\bm \sigma}_{ij}\times (-i {\bm \nabla}_{\bm r})] \cdot \hat{{\bm z}} - \mu~\delta_{{ij}} ~.
\ee
Here $\hat{\psi}^\dagger_{i}({\bm r})$ [$\hat{\psi}_{j}({\bm r})$] creates (destroys) an electron with real-spin label $i = \uparrow,\downarrow$ and band mass $m$, $\alpha$ measures the strength of Rashba SOC, 
${\bm {\sigma}} = (\sigma^1, \sigma^2)$ is a 2D vector of $2 \times 2$ Pauli matrices $\sigma^a$, $\mu$ is the chemical potential, and ${\bm {\hat z}}$ is a unit vector normal to the 2D plane where electrons are confined to move (the ${\hat {\bm x}} - {\hat {\bm y}}$ plane).  Diagonalization of $\hat{\cal H}_0$ yields two bands, $\xi_\lambda(k) = \kv^2/(2 m) +\lambda \alpha k - \mu$, $\lambda = \pm 1$ being the so-called ``chirality" index. Rashba SOC forces spins to lie on the ${\hat {\bm x}} - {\hat {\bm y}}$ plane and to be perpendicular to ${\bm k}$ at each point in momentum space [see Fig.~\ref{fig:one}a)].

The second term in the Hamiltonian $\hat{\cal H}$, $\hat{\cal H}_{\rm p}$, is an s-wave pairing Hamiltonian which is responsible for superconductivity: it physically corresponds to an attractive interaction of strength $-g$ with $g>0$, which is active only in a thin shell of momentum space around the Fermi surface.  The microscopic mechanism responsible for the appearance of the pairing term is not important here. 
The problem defined by $\hat{\cal H}_0 + \hat{\cal H}_{\rm p}$ has been studied by Gor'kov and Rashba~\cite{gorkov_prl_2001} who 
calculated the in-plane and out-of-plane spin susceptibilities $\chi_{\|(\perp)}(q = 0, \omega \to 0)$. Due to a mixture of spin-singlet and spin-triplet channels stemming from SOC, the GRSC develops a finite and anisotropic spin response. 

In this Letter we study the spin response of a GRSC at {\it finite} frequency $\omega$, taking into account also repulsive e-e interactions described by the last term in the Hamiltonian $\hat{\cal H}$, 
\be
\hat{\cal H}_{\rm e-e} = V \int d^2{\bm r}~{\hat \rho}_{\uparrow}({\bm r}) {\hat \rho}_{\downarrow}({\bm r})~,
\label{eeint}
\ee
where $V>0$ and the spin-resolved density operator is defined by ${\hat\rho}_i({\bm r}) = {\hat \psi}^\dagger_i({\bm r}) {\hat \psi}_i({\bm r})$.  
 We are interested in studying the collective dynamics of the system described by ${\hat {\cal H}}$ assuming that it remains in a phase characterized by a {\it hard} (finite in any direction of space) gap, despite the presence of repulsive e-e interactions. These lead to an effective reduction of the parameter $g$, in the spirit of the Anderson-Morel pseudopotential~\cite{Morel}.

We now  derive an effective low-energy action corresponding to the full Hamiltonian $\hat{\cal H}$ in terms of spin degrees-of-freedom only. The first step is to decouple the two quartic terms, $\hat{\cal H}_{\rm p}$ and $\hat{\cal H}_{\rm e-e}$, by means of a suitable Hubbard-Stratonovich (HS) transformation (see {\it e.g.} Refs.~\onlinecite{prange_prb_1979, benfatto_prb_2004}). For the pairing term $\hat{\cal H}_{\rm p}$ we introduce the complex HS field $\Delta_0({\bm r}, \tau)$, which describes the superconducting order parameter~\cite{benfatto_prb_2004}. We do the decoupling in the chiral basis: this allows us to work with Cooper pairs that are protected by time-reversal symmetry~\cite{gorkov_prl_2001}. Transforming back to the real-spin basis we get spin-triplet pairing in addition to the regular spin-singlet pairing~\cite{gorkov_prl_2001}. 

It is useful to rewrite $\hat{\cal H}_{\rm e-e}$ as~\cite{prange_prb_1979},
\be
\hat{\cal H}_{\rm e-e} = \frac{V}{4} \int d^2{\bm r}~\Bigg\{\hat{\rho}^{2}({\bm r}) - \Big[\sum_{a=1}^3\hat{s}_a({\bm r})\zeta_a\Big]^{2}\Bigg\}~,
\ee
where ${\hat\rho}({\bm r}) = \sum_i {\hat \rho}_i({\bm r})$ is the total-density operator, 
$\hat{s}_a({\bm r}) = \sum_{i,j} {{\hat \psi}}^{\dagger}_{i} ({\bm r}) \sigma^a_{ij}{{\hat \psi}}_{j} ({\bm r})$ is the usual spin-density operator, 
and ${\bm \zeta} = (\zeta_1,\zeta_2,\zeta_3)$ is an arbitrary unit vector in 3D space. To decouple  $\hat{\cal H}_{\rm e-e}$ by means of HS transformation we introduce four real HS fields~\cite{prange_prb_1979}: $\phi ({\bm r}, \tau)$ and ${\bm M}({\bm r}, \tau)$, which are conjugate to density fluctuations and spin fluctuations, respectively. 

The notation is considerably simplified by defining a four-component spinor $\hat{\Psi}^{\dagger}({\bm r}, \tau) = [\hat{\psi}^{\dagger}_{\uparrow} ~ \hat{\psi}^{\dagger}_{\downarrow} ~\hat{\psi}_{\uparrow} ~{\hat\psi}_{\downarrow} ]$ in real-spin space. The exact microscopic action corresponding to $\hat{\cal H}$ after the HS transformation can now be expressed in a compact form as (the variables ${\bm r}, \tau$ will be suppressed from now on when needed for brevity)
\ber\label{eq:exactaction}
{\cal S} = \int_0^\beta d\tau \int d^2{ \bm r} \Big[\frac{|\Delta_{0}|^{2}}{g} &+& \frac{\phi^{2} + {\bm M} \cdot {\bm M}}{V} \nonumber \\
&+& \bar{\Psi} \frac{(-G_{0}^{-1}+\Sigma_{0})}{2} \Psi \Big]~,
\eer
where $\beta = (k_{\rm B} T)^{-1}$, $\Sigma_{0}({\bm r}, \tau) = i \phi (\tau^3 \otimes \openone_{\sigma})$, and $\bar{\Psi}$ is the Grassmann variable corresponding to the fermionic field $\hat{\Psi}^{\dagger}$. 
Here $G_{0}^{-1}$ is the Green's function of the problem defined by $\hat{\cal H}_0 + \hat{\cal H}_{\rm p}$~\cite{gorkov_prl_2001} and is a $4 \times 4$ matrix given by 
\ber\label{eq:G0}
- G ^{-1}_{0} & =&   \partial_{t} \openone_{\tau} \otimes \openone_{\sigma}  + \tau^{3} \otimes {h}  + \alpha ~\{{\bm \Gamma}\times (-i {\bm \nabla}) \cdot \hat{{\bm z}}\} \nonumber \\ 
& &  + \frac{\tau^{1}+ i \tau^2}{2} \otimes {\bm\Delta} +  \frac{\tau^{1}- i \tau^2}{2} \otimes \bar{{\bm\Delta}}~.
\eer
The Pauli matrices $\tau^a$ act in the $2 \times 2$ Nambu-Gor'kov space and $\openone_\sigma$ ($\openone_\tau$) is the identity matrix in real-spin (Nambu-Gor'kov) space,
$\bm{\Gamma} = (\Gamma^1, \Gamma^2, \Gamma^3) \equiv (\tau^3 \otimes \sigma^1, \openone_\tau \otimes \sigma^2, \tau^3 \otimes \sigma^3)$ and ${\bm \Delta}$ is a $2 \times 2$ matrix whose diagonal (off-diagonal) elements are related to the triplet (singlet) order parameter [see Eq.~(\ref{eq:G0_full})].

At low energies, fluctuations of the amplitude of the order parameter $\Delta_{0}({\bm r}, \tau)$ do not play any role while phase fluctuations give rise to the Bogoliubov-Anderson mode~\cite{benfatto_prb_2004}.  To this end, we write $\Delta_{0} ({\bm r}, \tau) = \Delta e^{i \theta({\bm r}, \tau)}$, with $\Delta$ real. The amplitude $\Delta$ is fixed by the saddle-point equation $\delta {\cal S}/\delta \Delta = 0$, which yields the BCS equation~\cite{gorkov_prl_2001} 
[see Eq.~(\ref{eq: BCS1})].

The role of the phase field $\theta({\bm r}, \tau)$ can be made explicit in the action ${\cal S}$ by performing the following gauge transformation 
$\hat{\varphi}_{i} ({\bm r}, \tau) = \hat{\psi}_{i} ({\bm r}, \tau) e^{ i \theta({\bm r}, \tau)/2}$ to new fermionic fields $\hat{\varphi}_i({\bm r}, \tau)$. Writing the action ${\cal S}$  in terms of the new fermionic fields generates new self-energies in the round brackets in the second line of Eq.~(\ref{eq:exactaction}): $-G^{-1}_{0} + \Sigma_{0} \to  -G^{-1}_{0} + \Sigma$, where $\Sigma = \Sigma_1 + \Sigma_2 + \Sigma_3$ with
\ber
\Sigma_{1}({\bm r}, \tau) &=& \Bigg[i \Big(\frac{1}{2}\partial_\tau\theta + \phi\Big)  + \frac{(\nabla_{\bm r} \theta)^{2}}{8m} \Bigg] \tau^3 \otimes  \openone_{\sigma}  \nonumber \\ 
& - &
\frac{i}{2m}\left[\frac{{\bm \nabla^2_{\bm r}} \theta}{2} + ({\bm \nabla}_{\bm r} \theta) \cdot {\bm \nabla_{\bm r}}\right]  \openone_{\tau} \otimes \openone_{\sigma}~,
\eer
\be
\Sigma_{2}({\bm r}, \tau) = {\bm M} \cdot \bm {\Gamma} 
\;\; \mbox{and} \;\;
\Sigma_{3}({\bm r}, \tau) =  \frac{\alpha}{2} \left[{\bm \Gamma}\times ({\bm \nabla_{\bm r}} \theta) \right]\cdot {\hat{\bm z}}~. 
\ee

The fermionic part of the action can be integrated out (since it corresponds to a Gaussian functional integral for the partition function) leaving us with the following effective action
\ber \label{eq: S_eff2}
{\cal S}_{\rm eff} = \int_0^\beta d\tau \int d^2{ \bm r} &\Big[&\frac{\Delta^{2}}{g} + \frac{\phi^{2} + {\bm M} \cdot {\bm M}}{V}\Big] \nonumber \\
&- & \frac{1}{2} {\rm Tr}\left[\ln \left(-G_{0}^{-1} + \Sigma\right)\right]~,
\eer
where the symbol ``${\rm Tr}$" means a trace over all degrees of freedom (including space and imaginary time).   

To make further progress we need to expand the last term in ${\cal S}_{\rm eff}$ in powers of $\Sigma$. We keep terms up to second order  in the Fourier components of the fields $\phi_{\bm q},\theta_{\bm q}$ and  ${\bm M}_{\bm q}$.  A remarkable simplification occurs in the $q \to 0$ limit where the action reduces to the sum of  independent quadratic terms (see Appendix \ref{functionalintegral}).  Density and supercurrent oscillations on one hand and spin oscillations on the other hand  decouple.  As usual, the frequencies of collective modes are determined by the isolated poles of appropriate susceptibilities. For short range interactions, the density/current modes disperse linearly in $q$ and their frequency vanishes at $q=0$ as expected for a regular Goldstone mode. The spin modes, on the other hand, have a finite frequency, which increases with increasing $\Delta$ [consistent with the fact that the resistance of Cooper pairs to the twisting motion described in Fig.~\ref{fig:one}a) increases with increasing $\Delta$], but remains less than $2\Delta$, ensuring long lifetime.

{\it Collective spin oscillations. ---}  In the $q \to 0$ limit all the mixed response functions vanish  (see Appendix \ref{functionalintegral}) and the frequency of the collective spin mode $\omega_\|$ ($\omega_\perp$) at $q =0$ is given by the solution of the equation
\be\label{eq:collectivemodes}
2V^{-1} - \chi_{\|(\perp)}(0,\omega) = 0
\ee
with respect to $\omega$. In passing, we note that Eq.~(\ref{eq:collectivemodes}) can also be obtained diagrammatically from a vertex equation obtained by summing up ladder diagrams (see Appendix \ref{appendix: vertex}). 
In Eq.~(\ref{eq:collectivemodes}), $\chi_\|  = \chi_{\sigma^1\sigma^1} = \chi_{\sigma^2\sigma^2}$ 
and $\chi_\perp = \chi_{\sigma^3\sigma^3}$ are the in-plane and out-of-plane 
dynamical spin susceptibilities of the GRSC described by $\hat{\cal H}_0 + \hat{\cal H}_{\rm p}$, respectively. These are obtained from the analytical continuation, 
$i\nu_m \to \omega + i 0^+$, of the corresponding expressions in imaginary frequency:
\ber \label{eq:chibare}
 \chi_{\sigma^a\sigma^b}(0, i \nu_m)  &=&   - \frac{1}{2\beta A}\sum_{{\bm k}, n} {\rm Tr} \Big[\Gamma^a G_0({\bm k}, i \epsilon_n + i \nu_m/2) \nonumber\\
 &\times& \Gamma^b G_0({\bm k}, i \epsilon_n - i \nu_m/2) \Big]~,
\eer
where ``${\rm Tr}$" implies a trace over spin and Nambu-Gor'kov indices and $\nu_m$ ($\epsilon_n$) is a bosonic (fermionic) Matsubara frequency. 
After analytic continuation we find, at $T=0$,
\ber \label{eq:dynamicalbarechi}
\chi_\|(0, \omega) & = &   - \frac{1}{8 \pi} \int_{0}^{\infty} k dk 
\left( 1- \frac{ \xi_{+} \xi_{-} + \Delta^{2}}{E_{+}E_{-}}\right)   \nonumber \\
& \times & \left(\frac{1}{\omega + i 0^+ - {\cal E}} - \frac{1}{\omega + i 0^+ + {\cal E}} \right)
\eer
and $\chi_\perp(0,\omega) = 2 \chi_\|(0,\omega)$ [${\cal E} \equiv E_{+}(k) + E_{-}(k)$ and  $E^2_{\lambda}(k) \equiv \xi^2_{\lambda}(k)+\Delta^2$]. 
Due to the relation between out-of-plane and in-plane spin response functions, we will discuss only collective in-plane excitations.

\begin{figure}
\centering
\includegraphics[width=0.99\linewidth]{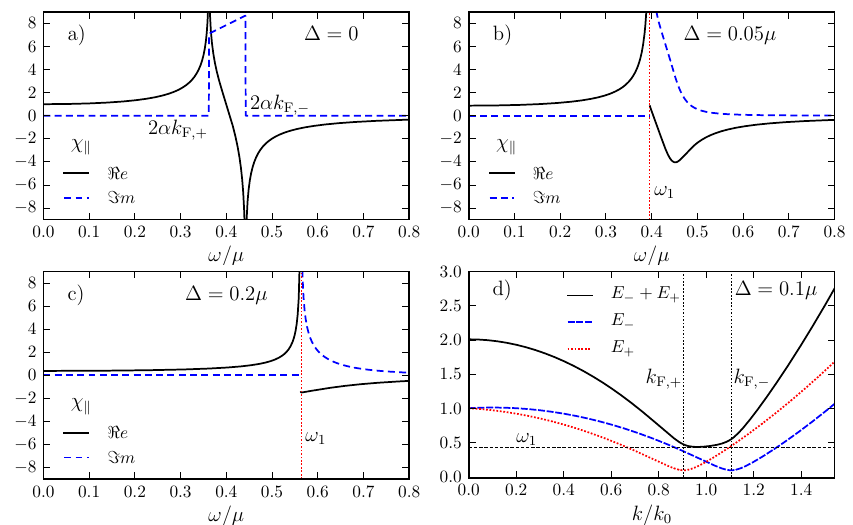}
\caption{(color online) Panel a) - c) The in-plane dynamical spin susceptibility $\chi_\|(0,\omega)$ [in units of the 2D density-of-states $m/(2 \pi)$] as a function of $\omega$ (in units of $\mu$) 
for increasing values of $\Delta$ (in units of $\mu$) and $V=0$.  The solid line represents $\Re e~\chi_{\|}(0,\omega)$, while the dashed line represents $\Im m~\chi_{\|}(0,\omega)$. 
Note that for finite $\Delta$,  $\Re e~\chi_{\|}(0,\omega)$ diverges at $\omega = \omega_1$ and that $\Im m~\chi_{\|}(0,\omega) = 0$ for $0< \omega < \omega_1$.
Panel d) The quantity $E_+(k) + E_-(k)$ as a function of $k$ (in units of $k_0$).  In this figure we have fixed $\alpha  = 0.2 \mu/k_0$ with $k_0 = \sqrt{2 m \mu}$. 
\label{fig:two}}
\end{figure}
%

We calculate $\chi_\|(0, \omega)$ numerically from Eq.~(\ref{eq:dynamicalbarechi}) and plot its real and imaginary parts in Fig.~\ref{fig:two}. 
In the limit $\Delta = 0$ ({\it i.e.} absence of superconductivity) -- see panel a) -- the imaginary part is non-zero only in the interval of frequencies 
between $2 \alpha k_{{\rm F}, +}$ and $2 \alpha k_{{\rm F}, -}$ [$k_{{\rm F}, \pm}$ being the minority (majority) Fermi wave vectors for the two Rashba bands $\xi_\lambda(k)$] 
and the real-part exhibits (logarithmic) singularities at these boundaries (see Appendix \ref{Appendix: normal}). When this result is inserted in Eq.~(\ref{eq:collectivemodes}), one finds a collective 
spin mode, which is undamped within this approximation. In a more refined theory (beyond Gaussian fluctuations), however, low-energy double electron-hole excitations damp this mode.
We now show that, at odds with the normal phase, in the superconducting state the mode lies (for a wide range of parameters) within the superconducting gap and thus cannot be damped by these excitations.

\begin{figure}
\centering
\includegraphics[width=0.99\linewidth]{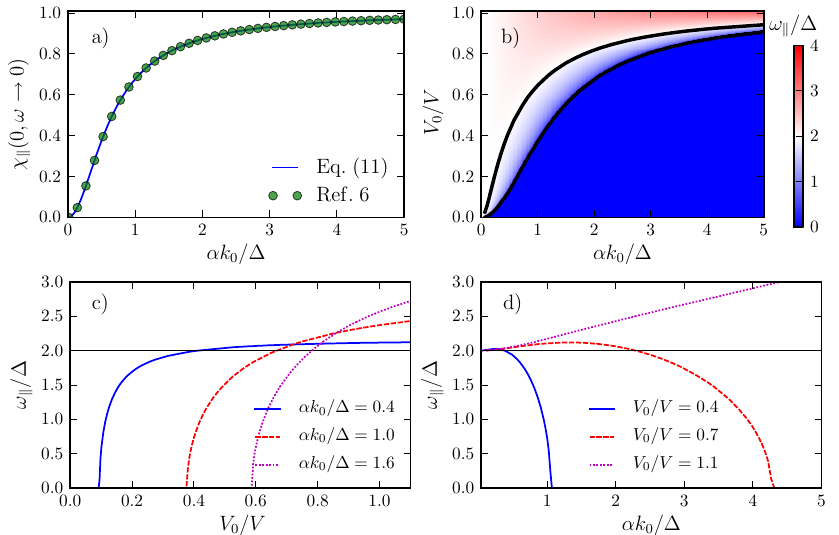}
\caption{(Color online) Panel a) The $\omega\to 0$ limit of $\chi_{\|}(0,\omega)$ as a function of $\alpha k_0/\Delta$. The solid line is the result obtained from Eq.~(\ref{eq:dynamicalbarechi}) while the filled circles are the result of Ref.~\onlinecite{gorkov_prl_2001}. Panel b) A 2D color plot of the frequency $\omega_{\parallel}$ of the in-plane collective spin mode (in units of $\Delta)$ as a function of the inverse of the strength of electron-electron repulsions ($V_0/V$, with $V_0 = 4 \pi/m$) and SOC ($\alpha k_0/\Delta$).  In this plot $\Delta = 0.1~\mu$. The top contour line is for $\omega_\| = 2 \Delta$ while the bottom contour line defines the boundary of the region in which $\omega_\| = 0$. The collective spin mode is undamped when it lies within the superconducting gap ($0< \omega_\| < 2 \Delta$), {\it i.e.} when $\omega_\|$ falls in the region enclosed by the two contour lines.  Panels c) and d) represents 1D cuts of the plot in panel b).\label{fig:three}}
\end{figure}

 In panels b) - c) we plot $\chi_\|(0, \omega)$ for finite $\Delta$.  In the superconducting state $\Re e~\chi_\|(0, \omega)$ exhibits a divergence at $\omega_1 \equiv {\rm min}_k [E_+(k) + E_-(k)]$. 
 In panel d) we plot $E_+(k) + E_-(k)$ as a function of $k$. In the region $0< \omega < \omega_1$, $\Im m~\chi_\|(0, \omega)$ is identically zero and, since $\Re e~\chi_\|(0,\omega)$ diverges for $\omega \to \omega_1$, there is always an in-plane collective spin mode with frequency $\omega_{\parallel} \approx \omega_{1}$ for weak repulsive interactions $V$. Our results for the frequency of the in-plane collective mode $\omega_\|$ as a function of $V$ and $\alpha$ (for a fixed value of $\Delta$) are summarized in Fig.~\ref{fig:three}.  Note that there is a wide range of parameters such that $\omega_\|$ lies within the superconducting gap, $0< \omega_\| < 2 \Delta$.  We also have checked that, as expected,  $\omega_\|$ increases with $\Delta$.

%

In summary, we have shown that quasi-two-dimensional superconductors with tunable spin-orbit coupling exhibit undamped collective spin oscillations  that can be excited by the application of a magnetic field or a supercurrent. The concerted action of spin-orbit coupling and electron-electron interaction is essential to the establishment of these collective oscillations.  
Since the frequency $\omega_\|$ of these oscillations is of the order of the superconducting gap $\Delta$ we expect that our findings might enable the realization of long-lived spin oscillators operating in the frequency range of $10~{\rm GHz}$  - $1~{\rm THz}$.
\acknowledgments 
We acknowledge financial support by the EU FP7 Programme under Grant Agreement No. 215368-SEMISPINNET (A.A. and M.P.), No. 234970-NANOCTM and No. 248629-SOLID (R.F), and by the NSF under Grant No. DMR-0705460 (G.V.). 
%

\begin{appendix}%
\section{The Green's function of a Gor'kov-Rashba superconductor}

Let us first consider the Green's function $G_{\rm c}$ of a Gor'kov-Rashba superconductor in the so-called ``chiral" basis. We remind the reader that we define ``Gor'kov-Rashba superconductor" the system defined by the Hamiltonian $\hat{\cal H}_0 + \hat{\cal H}_{\rm p}$, where the Rashba $\hat{\cal H}_0$ and pairing $\hat{\cal H}_{\rm p}$ Hamiltonians have been defined in the main text.
 
We start by defining the four-spinor $\hat{\Psi}^{\dagger}_{\rm c} ({\bm k}) = [\hat{\psi}^{\dagger}_{+}({\bm k})~\hat{\psi}_{+}(-{\bm k})~\hat{\psi}^{\dagger}_{-}({\bm k})~{\hat\psi}_{-}(-{\bm k}) ]$ in momentum space, where $\hat{\psi}_\lambda({\bm k})$ and $\hat{\psi}^{\dagger}_\lambda({\bm k})$ are field operators corresponding to the eigenstates of the Rashba Hamiltonian $\hat{\cal H}_0$ introduced in Eq.~(1) of the main text. In this ``chiral" basis the Matsubara Green's function corresponding to the Gor'kov-Rashba Hamiltonian $\hat{\cal H}_0 + \hat{\cal H}_{\rm p}$ [{\it i.e.} 
$G_{\rm c}({\bm k},\tau) = -\langle T_\tau \{ {\hat\Psi_{\rm c}} ({\bm k},\tau)  {\hat \Psi^{\dagger}_{\rm c}}({\bm k},0) \}\rangle$] is a $4 \times 4$ {\it block-diagonal} matrix (in Fourier transform with respect to imaginary time $\tau$):
\be
G_{\rm c}({\bm k}, i\epsilon_n) = \left(
\begin{array}{cc}
G_+({\bm k}, i\epsilon_n) & 0\\
0 & G_-({\bm k}, i\epsilon_n) 
\end{array}
\right)~,
\ee
where $G_\lambda({\bm k}, i\epsilon_n)$ are the following $2 \times 2$ matrices:
\begin{equation} 
\label{eq:greenfunction}
G_\lambda({\bm k}, i\epsilon_n) =  \frac{-i \epsilon_n \openone_{\tau} - \xi_{\lambda}(k)\tau^3
+ \lambda \Delta {\bm \tau}\cdot {\bm k}/k}{\epsilon_n^2 + E^2_{\lambda}(k)}~.
\end{equation}
Here $E^2_{\lambda}(k) \equiv \xi^2_{\lambda}(k)+\Delta^2$ and $\epsilon_n$ is a fermionic Matsubara frequency. The gap $\Delta$ of the Gor'kov-Rashba superconducting state is given by
\be
\Delta = \frac{g}{2}  \sum_{{\bm k}, \lambda} \lambda e^{i \phi_{\bm k}}~ \langle {\hat \psi}_{\lambda}(-{\bm k} ) {\hat \psi}_{\lambda} ({\bm k})\rangle~,
\ee
where $g > 0$ is the pairing coupling constant (see main text) and $\phi_{\bm k}$ is the angle between ${\bm k}$ and the ${\hat {\bm x}}$ axis.  
We emphasize that, following Ref.~\onlinecite{gorkov_prl_2001}, we have 
assumed that pairing occurs only between time-reversed partners within each Rashba spin-orbit-split band: 
$\langle {\hat \psi}_{\lambda}(-{\bm k} ) {\hat \psi}_{\bar{\lambda}} ({\bm k})\rangle = 0$ if ${\bar \lambda} = - \lambda$. 
The gap $\Delta$  is fixed by the saddle-point equation $\delta {\cal S}/\delta \Delta = 0$, [${\cal S}$ is given in Eq.~(4) of the main text]  which yields the mean-field BCS equation~\cite{gorkov_prl_2001}
\be \label{eq: BCS1}
1= \frac{g}{2}\frac{1}{A}\sum_{\kv, \lambda} \frac{\tanh{[\beta E_{\lambda}(k)/2]}}{2E_{\lambda}(k)}~,
\ee
where $A$ is the system's area. 

The Green's function $G_0$ introduced in Eq.~(5) of the main text is explicitly given by the following $4 \times 4$ matrix:
 \begin{widetext}
 \be \label{eq:G0_full}
G^{-1}_{0}({\bm r}, \tau) = -
\begin{pmatrix} 
\partial_{\tau} - \nabla^{2}_{\bm r}/(2m)  &  \alpha (\partial_{x} - i\partial_{y})&  {\bm \Delta}_{11}& {\bm \Delta}_{12}\\
-\alpha (\partial_{x} + i\partial_{y}) & \partial_{\tau}  - \nabla^{2}_{\bm r}/(2m)  &  -{\bm \Delta}_{12} & {\bm \Delta}_{22} \\
\bar{\bm \Delta}_{11}& \bar {\bm \Delta}_{12}& \partial_{\tau} + \nabla^{2}_{\bm r}/(2m)  & \alpha (\partial_{x} + i\partial_{y}) \\
-\bar{\bm \Delta}_{12}& \bar{\bm \Delta}_{22}& - \alpha (\partial_{x} - i\partial_{y}) &\partial_{\tau}+  \nabla^{2}_{\bm r}/(2m)    
\end{pmatrix}~,
\ee
\end{widetext}
where 
${\bm \Delta}_{11}({\bm r}, \tau) = \langle {\hat \psi}_{\uparrow}({\bm r}, \tau) {\hat \psi}_{\uparrow}({\bm r}, \tau)\rangle$ and ${\bm \Delta}_{22} ({\bm r}) = \langle {\hat \psi}_{\downarrow}({\bm r}, \tau) {\hat \psi}_{\downarrow}({\bm r}, \tau)\rangle$ are related to the triplet order parameter (which, of course, arises only because of the presence of spin-orbit coupling) and ${\bm \Delta}_{12}({\bm r}, \tau) = \langle {\hat \psi}_{\downarrow}({\bm r}, \tau) {\hat \psi}_{\uparrow}({\bm r}, \tau)\rangle$ is the singlet order parameter. 

The $4 \times 4$ Green's function $G_0$ in the real-spin basis can be related to the Green's function $G_{\rm c}$ in the chiral basis: we find (in Fourier transform with respect to space and imaginary time)
 \be \label{eq:GF_spin1}
G_{0}^{11} ({\bm k},i\epsilon_{n}) = G_{\rm s} (k,i\epsilon_{n}) \openone_{\sigma} + G_{\rm a}(k,i\epsilon_{n}) \left[\left({\hat {\bm k}} \times {\bm \sigma}\right) \cdot {\hat{\bm z}}\right]~, ~
\ee
\be
G_{0}^{12}({\bm k},i\epsilon_{n}) = F_{\rm s}(k,i\epsilon_{n}) \sigma^2+  F_{\rm a} (k,i\epsilon_{n})\left[\left({\hat {\bm k}} \times {\bm \sigma}\right) \cdot {\hat{\bm z}}\right] \sigma^2~, 
\ee
\be
G_{0}^{21} ({\bm k},i\epsilon_{n}) = F_{\rm s}(k,i\epsilon_{n}) \sigma^2 - F_{\rm a} (k,i\epsilon_{n})  \left[\left({\hat {\bm k}} \times {\bm \sigma}\right) \cdot {\hat{\bm z}}\right]^{\rm T}\sigma^2~,
\ee
and 
\be \label{eq:GF_spin4}
G_{0}^{22} ({\bm k},i\epsilon_{n}) = {\widetilde G}_{\rm s}(k,i\epsilon_{n}) \openone_{\sigma}- {\widetilde G}_{\rm a}(k,i\epsilon_{n}) \left[\left({\hat {\bm k}} \times {\bm \sigma}\right) \cdot {\hat{\bm z}}\right]^{\rm T}~.
\ee 
Here ${\hat{\bm k}} = {\bm k}/|{\bm k}|$  and we have defined 
\ber \label{eq:Green_sum}
&&2G_{\rm s/a} (k,i\epsilon_{n}) \equiv  G^{11}_{+}  (k,i\epsilon_{n})\pm G^{11}_{-} (k,i\epsilon_{n}) \\
&& \frac{-i \epsilon_{n} - \xi_+ (k)}{\epsilon_{n}^{2}+ \xi_+ (k)^{2} + \Delta^2} \pm \frac{-i \epsilon_{n} - \xi_- (k)}{\epsilon_{n}^{2}+ \xi_- (k)^{2} + \Delta^2}~, \nonumber
\eer
\ber
&&2 {\widetilde G}_{\rm s/a} (k,i\epsilon_{n}) \equiv  G^{22}_{+}  (k,i\epsilon_{n}) \pm G^{22}_{-}(k,i\epsilon_{n}) \\
&&= \frac{-i \epsilon_{n} + \xi_+ (k)}{\epsilon_{n}^{2}+ \xi_+ (k)^{2} + \Delta^2} \pm \frac{-i \epsilon_{n} + \xi_- (k)}{\epsilon_{n}^{2}+ \xi_- (k)^{2} + \Delta^2}~, \nonumber 
\eer
and 
\ber \label{eq:Green_sum2}
& & 2 F_{\rm s/a}(k,i\epsilon_{n})\equiv  e^{i \phi_{{\bm k}}}\left[G^{12}_{+} (k,i\epsilon_{n}) \mp G^{12}_{-} (k,i\epsilon_{n}) \right] \\
&& =   \Delta \left( \frac{1}{\epsilon_{n}^{2}+ \xi_+ (k)^{2} + \Delta^2} \pm \frac{1}{\epsilon_{n}^{2}+ \xi_- (k)^{2} + \Delta^2}\right). \ \ \ \ \nonumber
\eer
In Eqs.~(\ref{eq:GF_spin1})-(\ref{eq:Green_sum2}) $G^{\alpha\beta}_0$ ($G^{\alpha\beta}_\pm$) are the elements of the matrix $G_0$ ($G_\pm$).

\section{The effective action, Gaussian fluctuations, and spin collective excitations in the long-wavelength limit}
\label{functionalintegral}
To include  the phase fluctuations of the order parameter in our study (ignoring the fluctuations in the modulus  of the order parameter), we perform a gauge transformation to new fermionic fields
\be \label{eq:varphi1}
{\hat{\varphi}}^{\dagger}_{i} ({\bm r}, \tau) = {\hat{\psi}}^{\dagger}_{i} ({\bm r}, \tau) e^{ -i \theta({\bm r}, \tau)/2}~,
\ee
and
\be \label{eq:varphi2}
{\hat{\varphi}}_{i} ({\bm r}, \tau) = {\hat{\psi}}_{i} ({\bm r}, \tau) e^{ i \theta({\bm r}, \tau)/2}~.
\ee
Using Eqs.~(\ref{eq:varphi1})-(\ref{eq:varphi2}) in the definition of the action ${\cal S}$ given in Eq.~(4) of the main text we find:
\ber\label{eq:exactaction2}
{\cal S} &=& \int_0^\beta d\tau \int d^2{ \bm r}  \Bigg[\frac{\Delta^{2}}{g} + \frac{\phi^{2} + {\bm M} \cdot {\bm M}}{V} \nonumber \\
&+&  \bar{\varPhi} \frac{(-G_{0}^{-1}+\Sigma_{1} +\Sigma_{2} +\Sigma_{3} )}{2} \varPhi \Bigg]~,
\eer
where $\varPhi ({\bm r}, \tau) = [\hat{\varphi}^{\dagger}_{\uparrow} ~ \hat{\varphi}^{\dagger}_{\downarrow} ~\hat{\varphi}_{\uparrow} ~{\hat\varphi}_{\downarrow} ]$ is a four component spinor, and $\Sigma_1$, $\Sigma_2$ and $\Sigma_3$ are given by Eqs. (6) - (8) of the main text. Now the fermionic fields ($\varPhi$) can be integrated out of the partition function, which is of the form $Z = Z_0 \int D[\varPhi] \exp{\left({\bar \varPhi} B \varPhi\right)/2}$,  by performing a Gaussian integral over the Grassman variables. We use the following relations for an arbitrary matrix $B$,
\ber \label{eq: integration}
 &&\int D[\varPhi] \exp{\left(\frac{1}{2}{\bar \varPhi} B \varPhi\right)} = \sqrt{{\rm Det}[B]} \nonumber \\ 
&&=   \exp{\left[\frac{1}{2}\ln ({\rm Det}[B])\right]} =\exp \left(\frac{1}{2}{\rm Tr}\ln[B]\right)~.~~
\eer
Using Eq.~(\ref{eq: integration}) in Eq.~(\ref{eq:exactaction2}) immediately gives the effective action ${\cal S}_{\rm eff}$ reported in Eq.~(\ref{eq: S_eff2}) of the main text.

To expand ${\cal S}_{\rm eff}$ up to second order in $\Sigma$, we use the identity
\ber
{\rm Tr}\left[\ln \left( -G_{0}^{-1} + \Sigma\right)\right] &=&{\rm Tr}\left[\ln \left(-G_{0}^{-1}\right)\right] \nonumber \\
&-&{\rm Tr}\left[\sum_{n=1}^{\infty}\frac{1}{n}(G_{0}\Sigma)^{n}\right]~.~~
\eer
Using Eq.~(8) in the main text and the identity above we find, up to second order in $\Sigma$,
\ber
 {\cal S}_{\rm eff} &\approx& \int d\tau d{ \bm r} \left[ \frac{\Delta^{2}}{g} + \frac{\phi^{2}}{V} + \frac{{\bm M} \cdot {\bm M}}{V}  \right] \nonumber \\
&  -& \frac{1}{2}{\rm Tr}\left[\ln \left(-G_{0}^{-1}\right)\right]  + \frac{1}{2}{\rm Tr}[G_{0}\Sigma] + \frac{1}{4} {\rm Tr}[G_{0}\Sigma G_{0} \Sigma] \nonumber \\ 
&+& {\cal O}(\Sigma^3)~,
\eer
where the trace ``${\rm Tr}$" is taken over all degrees of freedom including space and imaginary time.  Note that 
in the last term of the previous equation we have to keep only terms up to second order in the fluctuating fields $\theta$, $\phi$, and ${\bm M}$ (Gaussian fluctuations).
 
We can rewrite ${\cal S}_{\rm eff}$ as
\be
{\cal S}_{\rm eff} \approx \int d\tau d{ \bm r} \frac{\Delta^{2}}{g} -{\rm Tr}\left[\ln \left(-G_{0}^{-1}\right)\right] + {\cal S}_{\rm fluct}^{(2)}~,
\ee
where ${\cal S}_{\rm fluct}^{(2)}$ is the second-order contribution to ${\cal S}_{\rm eff}$ due to the fluctuating fields. Using Fourier transforms with respect to space and imaginary time 
and defining  $\Pi({\bm q},i\nu_{m}) = [\theta ({\qv},i\nu_{m})~ \phi ({\qv},i\nu_{m})~ {\bm M}({\bm q},i\nu_{m})]$ we find
\be \label{eq:action2}
{\cal S}_{\rm fluct}^{(2)} = \sum_{{\bm q}, m}\Pi ({\bm q},i\nu_{m}) ~ N({\bm q}, i \nu_{m}) ~\Pi^{\rm T}(-{\bm q},-i\nu_{m})~,
\ee
where $N$ is a $5 \times 5$ matrix whose elements are given by various dynamical response functions and $\nu_m$ is a bosonic Matsubara frequency. Different elements of the matrix $N(\bm q, i \nu_{m})$ are given by 
\ber\label{eqstart}
& &N_{11} (\bm q, i \nu_{m}) = \frac{ \nu_{m}^{2}}{8} \chi_{\rho \rho}(\bm q, i \nu_{m}) 
+ \frac{1}{8}  \sum_{i,j}  q_{i} q_{j} \Big[ \delta_{ij} \frac{ \rho_{\rm mf}}{m} \nonumber \\
& &- \chi_{j^{i}j^{j}} (\bm q, i \nu_{m})\Big] - \frac{i }{8}  \nu_{m}\sum_{i} q_{i}\big[\chi_{j^{i}\rho}(\bm q, i \nu_{m}) \nonumber \\
& &+  \chi_{j^{i}\rho}(-{\bm q},-i \nu_{m})\big] +  \frac{\alpha^{2}}{8} q^{2} \chi_{\phi \phi} (\bm q, i \nu_{m}) \nonumber \\ 
& & + \frac{\alpha }{8}   \big[\chi_{\phi \rho}(\bm q, i \nu_{m}) - \chi_{\phi \rho}(-\bm q, -i \nu_{m}) \big] \nonumber \\ 
&& - \frac{i \alpha}{8} q \sum_{i} q_{i}  \big[ \chi_{\phi j^{i}}(\bm q, i \nu_{m})  + \chi_{\phi j^{i}}(-\bm q, -i \nu_{m})  \big]~,~~
\eer
\be
N_{22}(\bm q, i \nu_{m}) =   \frac{1}{V}  + \frac{1}{2} \chi_{\rho \rho}(\bm q, i \nu_{m})~,
\ee
\ber
& &N_{12} (\bm q, i \nu_{m}) =  -\frac{i}{4}  \nu_{m}  \chi_{\rho \rho}(\bm q, i \nu_{m})   \\ 
& & - \frac{1}{4}  \sum_{i} q_{i}\chi_{j^{i}\rho}(-\bm q, -i \nu_{m}) + \frac{i \alpha}{4} q   \chi_{\phi \rho}(-\bm q, -i \nu_{m})~, \nonumber 
\eer
\ber
& &N_{21} (\bm q, i \nu_{m})= \frac{i}{4}  \nu_{m}  \chi_{\rho \rho}(\bm q, i \nu_{m}) \nonumber \\
& & +\frac{1}{4}  \sum_{i} q_{i}\chi_{j^{i}\rho}(\bm q, i \nu_{m}) + \frac{i \alpha}{4} q   \chi_{\phi \rho}(\bm q, i \nu_{m})~, ~~
\eer
\be
N_{ab} (\bm q, i \nu_{m})=   \frac{1}{V} \delta_{ab} - \frac{1}{2}\chi_{\sigma^{a-2} \sigma^{b-2}}(\bm q, i \nu_{m})~,
\ee
with $a, b \in [3,4,5]$,
\ber
&&N_{1a}  (\bm q, i \nu_{m})= - \frac{1}{4}  \nu_{m} \chi_{\sigma^{a} \rho}(\bm q, i \nu_{m}) \nonumber \\
&&+ \frac{i }{4} \sum_{i} q_{i}  \chi_{\sigma^{a} j^{i}}(\bm q, i \nu_{m}) 
- \frac{ \alpha}{4} q \chi_{\sigma^{a}\phi} (\bm q, i \nu_{m})~,~~
\eer
\ber
&&N_{a1} (\bm q, i \nu_{m}) = \frac{1}{4}  \nu_{m} \chi_{\sigma^{a-2} \rho}(-\bm q, -i \nu_{m}) \\
&&- \frac{i}{4} \sum_{i} q_{i}  \chi_{\sigma^{a-2} j^{i}}(-\bm q, -i \nu_{m})- \frac{ \alpha}{4} q \chi_{\sigma^{a-2}\phi} (-\bm q, -i \nu_{m})~,\nonumber \eer
\be
N_{2a} (\bm q, i \nu_{m}) = - i  \chi_{\sigma^{a-2} \rho}(\bm q, i \nu_{m})/2~,
\ee
and, finally,
\be\label{eqend}
N_{a2} (\bm q, i \nu_{m}) =- i  \chi_{\sigma^{a-2} \rho}(-\bm q, -i \nu_{m})/2~.
\ee
The quantity $\rho_{\rm mf}$ which appears in Eq.~(\ref{eqstart}) is given by 
\be 
\rho_{\rm mf}= \frac{1}{2\beta}\sum_{\bm k, n} {\rm Tr}[G_0({\bm k},\epsilon_{n}) \tau^3 \otimes \openone_{\sigma}]~,
\ee
and physically corresponds to the total electron density (superfluid and normal component)~\cite{benfatto_prb_2004}.

The response functions $\chi$ that appear in Eqs.~(\ref{eqstart})-(\ref{eqend}) are dynamical susceptibilities of the Gor'kov-Rashba superconducting state in the absence of electron-electron interactions:
\begin{widetext}
\be\label{chibegin}
 \chi_{\rho \rho}(\bm q, i \nu_{m}) = -\frac{1}{2\beta} \sum_{\bm k, n} {\rm Tr}
\left[G_0({\bm k}^{+}, \epsilon_{n}^{+})~\tau^3\otimes\openone_{\sigma}~G_0({\bm k}^{-}, \epsilon_{n}^{-})~\tau^3\otimes\openone_{\sigma}~ \right]~, 
\ee
\be
\chi_{j^{i} j^{j}}(\bm q, i \nu_{m}) = -\frac{1}{2\beta} \sum_{\bm k, n} \frac{k_ik_j}{m^{2}}~ {\rm Tr}
\left[G_0({\bm k}^{+}, \epsilon_{n}^{+})~\openone_{\tau} \otimes \openone_{\sigma}~G_0({\bm k}^{-}, \epsilon_{n}^{-})~\openone_{\tau} \otimes \openone_{\sigma}~   \right]~, 
\ee
\be
\chi_{j^{i} \rho}(\bm q, i \nu_{m}) = -\frac{1}{2\beta} \sum_{\bm k, n} \frac{k_{i}}{m}~{\rm Tr}
\left[G_0({\bm k}^{+}, \epsilon_{n}^{+})~\tau^3\otimes\openone_{\sigma}~G_0({\bm k}^{-}, \epsilon_{n}^{-})~\openone_{\tau} \otimes \openone_{\sigma}~   \right] ~,
\ee
\be\label{eq:spinspin}
\chi_{\sigma^{i} \sigma^{j}}(\bm q, i \nu_{m}) = -\frac{1}{2\beta} \sum_{\bm k, n} {\rm Tr}\left[G_0({\bm k}^{+}, \epsilon_{n}^{+}) \Gamma^{j}~G_0({\bm k}^{-}, \epsilon_{n}^{-})\Gamma^{i}  \right]~,
\ee
\be
 \chi_{\sigma^{i} \rho}(\bm q, i \nu_{m}) = -\frac{1}{2\beta} \sum_{\bm k, n} {\rm Tr}
\left[G_0({\bm k}^{+}, \epsilon_{n}^{+})~\tau^3\otimes \openone_{\sigma}~G_0({\bm k}^{-}, \epsilon_{n}^{-})\Gamma^{i} \right] ~,
\ee
\be
\chi_{\sigma^{i} j^{j}}(\bm q, i \nu_{m}) = -\frac{1}{2\beta} \sum_{\bm k, n}\frac{ k_j}{m}~ {\rm Tr}
\left[G_0({\bm k}^{+}, \epsilon_{n}^{+})~\openone_{\tau} \otimes \openone_{\sigma}~G_0({\bm k}^{-}, \epsilon_{n}^{-})\Gamma^{i} \right] ~,
\ee
\be
\chi_{\phi \phi} (\bm q, i \nu_{m}) =  -\frac{1}{2\beta} \sum_{\bm k, n}{\rm Tr}\left[G_0({\bm k}^{+}, \epsilon_{n}^{+})\Gamma_{\phi}~G_0({\bm k}^{-}, \epsilon_{n}^{-})\Gamma_{\phi} \right]~, 
\ee
\be \chi_{\sigma^{i} \phi} (\bm q, i \nu_{m}) =  -\frac{1}{2\beta} \sum_{\bm k, n}{\rm Tr}\left[G_0({\bm k}^{+}, \epsilon_{n}^{+})\Gamma_{\phi}~G_0({\bm k}^{-}, \epsilon_{n}^{-})\Gamma^{i} \right]~, 
\ee
\be
\chi_{\phi \rho} (\bm q, i \nu_{m}) =  -\frac{1}{2\beta} \sum_{\bm k, n}{\rm Tr}\left[G_0({\bm k}^{+}, \epsilon_{n}^{+})~\tau^3\otimes\openone_{\sigma}~G_0({\bm k}^{-}, \epsilon_{n}^{-})~\Gamma_{\phi}\right]~,
\ee
and, finally,
\be\label{chiend}
\chi_{\phi j^{i}} (\bm q, i \nu_{m}) =  -\frac{1}{2\beta} \sum_{\bm k, n}  \frac{ k_i}{m}~{\rm Tr}\left[G_0({\bm k}^{+}, \epsilon_{n}^{+})~\openone_{\tau} \otimes \openone_{\sigma}~G_0({\bm k}^{-}, \epsilon_{n}^{-})\Gamma_{\phi} \right]~.
\ee
\end{widetext}
In Eqs.~(\ref{chibegin})-(\ref{chiend}) ${\bm k}^{\pm} = {\bm k} \pm {\bm q}/2$, $\epsilon_n^{\pm} = \epsilon_n \pm \nu_m/2$, the trace ``${\rm Tr}$" is to be taken over the Nambu-Gor'kov and spin indices,  and the $4 \times 4$ matrix $\Gamma_{\phi}$ is 
\ber
\Gamma_{\phi} ({\bm q}) =  \begin{pmatrix} 
0 &  e^{i \phi_{\bm q}}& 0 & 0 \\
- e^{-i \phi_{\bm q}}& 0 & 0 & 0 \\
0& 0 & 0 &  e^{-i \phi_{\bm q}} \\
0& 0&  -e^{i \phi_{\bm q}}& 0
\end{pmatrix}~.
\eer
The matrices $\Gamma^i$ with $i = 1 \dots 3$ have been introduced in the main text. Notice that the $4 \times 4$ matrix $\tau^3 \otimes \openone_\sigma$ corresponds to the density operator, the $4 \times 4$ matrices $\Gamma^i$ correspond to the spin operators $\hat{s}_i$, and the $4 \times 4$ matrix $(\openone_\tau \otimes \openone_\sigma) k_i /m$ corresponds to the current operator $\hat{j}_i$. 

In deriving Eq.~(\ref{eq:action2}) we have used the following relations: $\chi_{AB}({\bm q}, i \nu_{m}) = \chi_{BA}({-\bm q}, -i \nu_{m})$ where $A$ or $B$ denote 
density, spin or current operators and $\chi_{\phi A}({\bm q}, i \nu_{m}) = - \chi_{A \phi}(-{\bm q}, -i \nu_{m})$.
Note that for an ordinary superconductor in the absence of spin-orbit-coupling ($\alpha = 0$), the quantities $N_{11}, N_{12},N_{21},N_{22}$ reduce to the elements of the matrix ${\hat B}(q)$ defined 
in Appendix A of Ref.~\onlinecite{benfatto_prb_2004} and determine the frequency of the Bogoliubov-Anderson mode once the limit $q \to 0$ is taken.

To find the energy-momentum dispersion of the collective spin excitations, we obtain an effective action in terms of the spin degrees-of-freedom ${\bm M}$ only by performing a Gaussian integral over the fields $\phi$ and $\theta$ in Eq.~(\ref{eq:action2}). The ``spin-only" action ${\cal S}^{(2)}_{\bm M}$ can be expressed in a compact form, by defining $\Pi^{1}({\bm q},i\nu_{m} )   = [M_{1}({\bm q},i\nu_{m})~M_{2}({\bm q},i\nu_{m})~M_{3}({\bm q},i\nu_{m}) ]$, as
\be
{\cal S}^{(2)}_{\bm M}= \sum_{\bm q, m}\Pi^{1} ({\bm q},i\nu_{m})~Q({\bm q}, i\nu_{m})~\Pi^{1 {\rm T}}(-{\bm q},-i\nu_{m})~,\\
\ee
where $Q$ is a $3 \times 3$ matrix whose elements are given  by 
\begin{widetext}
\be \label{eq:Q}
Q_{cd}(\bm q, i\nu_{m}) =  N_{c+2 ~d+2} + A^{-1} \left[ N_{1~d+2} ( N_{22} N_{c+2~ 1} - N_{21} N_{c+2~2}) + N_{2~d+2}(N_{11} N_{c+2~2} - N_{12} N_{c+2~1}) \right] ~,
\ee
\end{widetext}
with $c,d \in [1,2,3]$ and where $ A (\bm q, i\nu_{m}) = N_{11} N_{22} - N_{21} N_{12}$.
Note that the first term in $Q_{cd}$ originates from the correction to the bare dynamical response function due to electron-electron interactions, while the second term originates from 
the coupling of spin fluctuations to phase fluctuations. 
The collective spin modes of the system can be found by solving
\be\label{eqcollectivemodes}
{\rm det}[Q({\bm q}, i\nu_m \to \omega + i 0^+)] = 0~.
\ee

As mentioned in the main text, in this work we are interested in finding the frequency of the collective spin modes in the long-wavelength $q \to 0$ limit. In this limit Eq.~(\ref{eq:action2}) simplifies considerably and the effective action decouples into separate terms corresponding to  supercurrent/density oscillations and spin oscillations, respectively. More specifically in the $q \to 0$ limit we have,
\ber \label{eq: chi_cross1}
&&\chi_{\sigma^2 \rho}(0, i\nu_{m})  = \chi_{\sigma^1 \rho}(0, i\nu_{m}) = \chi_{\sigma^2 \sigma^3}(0,i\nu_{m})  \nonumber \\
&&= \chi_{\sigma^1 \sigma^3}(0,i\nu_{m}) = 0~.
\eer
Moreover, 
\be  \label{eq: chi_cross2}
\chi_{\sigma^1 \sigma^2} ({\bm q},i\nu_{m}) = \chi_{\sigma^3 \rho}({\bm q},i\nu_{m}) = 0~
\ee
for every finite ${\bm q}$. In the limit $q\to 0$ Eqs.~(\ref{eq: chi_cross1})-(\ref{eq: chi_cross2}) give
\ber \label{eq: N}
&&N_{13}=N_{14}=N_{15} =N_{31} =N_{41} =N_{51} =N_{23} \nonumber \\
&& = N_{24}=N_{25} =N_{32} =N_{42} =N_{52} =0~, 
\eer
and
\be \label{eq: N2}
N_{34} = N_{35} = N_{43} = N_{45} = N_{53} = N_{35} =0~.
\ee
Using Eq.~(\ref{eq: N}) in Eq.~(\ref{eq:Q}) we obtain 
\be 
Q_{cd}(0, i\nu_{m}) = N_{c+2~d+2}(0,i\nu_{m})~.
\ee
In other words,  phase/density fluctuations do {\it not} couple to spin fluctuations in the long-wavelength limit. 
Using Eq.~(\ref{eq: N2}) in Eq.~(\ref{eq:Q}) we obtain $Q_{cd}= 0$ for $c\neq d$, {\it i.e.}, all the off-diagonal components of the matrix $Q$ are zero in the long wavelength limit. Eq.~(\ref{eqcollectivemodes}) thus reduces to
\ber\label{eqcollectivemodeszeroq}
&&{\rm det}[Q(0, i\nu_m \to \omega + i 0^+)] = \left[\frac{2}{V} - \chi_{\sigma^1\sigma^1}(0, \omega)\right] \nonumber \\
&&\times \left[\frac{2}{V} - \chi_{\sigma^2\sigma^2}(0, \omega)\right]\left[\frac{2}{V} - \chi_{\sigma^3\sigma^3}(0, \omega)\right] =0~,~~~~~
\eer
with $\chi_{\sigma^1\sigma^1}(0, \omega) = \chi_{\sigma^2\sigma^2}(0, \omega)$.

Note that we can also obtain Eq.~(\ref{eqcollectivemodeszeroq}) directly from ${\cal S}_{\rm fluct}^{(2)}$. Equation~(\ref{eq: N}) implies that the matrix $N$ in Eq.~(\ref{eq:action2}) has a block diagonal form comprising an upper $2 \times 2$ block corresponding to $\theta - \phi$ fields and a lower $3 \times 3$ block corresponding to the ${\bm M}$ fields.
Thus the $q = 0$ component of the action ${\cal S}_{\rm fluct}^{(2)}$ can be expressed as a product of a``phase only" action and a``spin only" action, {\it i.e.} 
\be \label{eq: decouple}
\left. {\cal S}_{\rm fluct}^{(2)}\right|_{q =0}= \left. {\cal S}^{(2)}_{\theta, \phi}\right|_{q =0} \times \left. {\cal S}^{(2)}_{\bm M}\right|_{q =0}~.
\ee
Moreover, Eq.~(\ref{eq: N2}) implies that the  lower $3 \times 3$ block of the $N$ matrix in Eq.~(\ref{eq:action2}) is diagonal in the long wavelength limit. We thus obtain
\ber
\left. {\cal S}^{(2)}_{\bm M}\right|_{q =0} = \frac{1}{2}\sum_{\nu_{m}, i} & & M_{i}(0, i \nu_m) \left[\frac{2}{V} - \chi_{\sigma^{i}\sigma^{i}}(0,i\nu_{m}) \right] \nonumber \\
& & \times M_{i}(0, i \nu_m)~,
\eer
which gives the same condition for the existence of collective spin modes as Eq.~(\ref{eqcollectivemodeszeroq}).
\section{The ladder sum and the vertex equation}
\label{appendix: vertex}

In this Section we show that the equation
\ber\label{polesexplicit}
&& \frac{2}{V} - \chi_{\sigma^1\sigma^1}(0, \omega) = \frac{2}{V} - \chi_{\sigma^2\sigma^2}(0, \omega) \nonumber \\
&& = \frac{2}{V} - \chi_{\sigma^3\sigma^3}(0, \omega) = 0
\eer 
for the collective spin excitations that we found in the previous section, Eq.~(\ref{eqcollectivemodeszeroq}), can also be obtained diagrammatically.

In the (conserving) ladder approximation~\cite{Schrieffer, Nambu} the dynamical spin response function ${\widetilde \chi}_{\sigma^2\sigma^2}$ in the presence of electron-electron interactions is given by
\ber \label{eq:chi}
\widetilde{\chi}_{\sigma^2\sigma^2}(\qv,i \nu_{m}) &=& - \frac{ 1}{2 \beta}
\sum_{\bm k,n}{\rm Tr} \Big[\Gamma^{2} G_0({\bm k}^+, i\epsilon^+_{n})  \Lambda (\qv,  i\nu_{m}) \nonumber \\
&\times&G_0({\bm k}^-, i\epsilon^-_{n}) \Big]~.
\eer
The vertex function $\Lambda(\qv,  i\nu_{m})$ is  a ``dressed" version of the bare vertex $\Gamma^{2} = {\openone_{\tau} \otimes \sigma^2} $ 
and it accounts for the interplay between electron-electron interactions and the external electromagnetic field. Similar equations hold for ${\widetilde \chi}_{\sigma^1\sigma^1}$ and ${\widetilde \chi}_{\sigma^3\sigma^3}$.
%
\begin{figure}[t]
\begin{center}
\includegraphics[width=1.04 \linewidth]{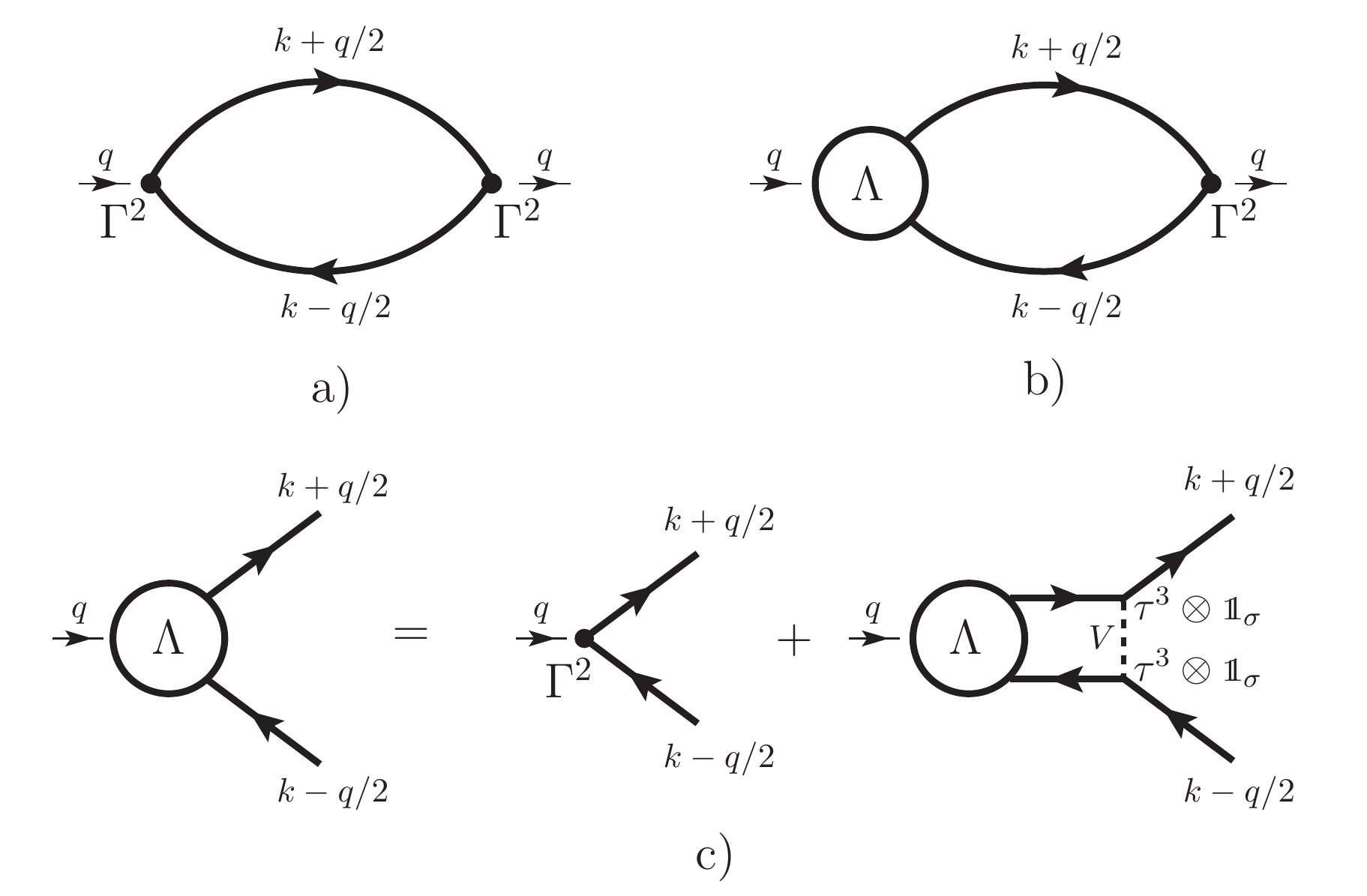}
\caption{Panel a)  The bare bubble $\chi_{\sigma^2\sigma^2}$ constructed with the Gor'kov-Rashba Green's function $G_0$ [see Eq.~(\ref{eq:spinspin})].  Panel b) The dynamical response function 
${\widetilde \chi}_{\sigma^2\sigma^2}$ with the inclusion of vertex corrections due to electron-electron interactions. Panel c) The vertex equation in the ladder approximation. 
In this figure, $k\pm q/2$ denotes the 4 vector ($\epsilon_{n}\pm \nu_m/2, {\bm k} \pm {\bm q}/2$) and $\Gamma^{2} = {\openone_{\tau} \otimes \sigma^2}$ denotes the spin operator $\hat{s}_2$. \label{fig:ladder}}
\end{center}
\end{figure}
%

The vertex $\Lambda$  is a $4\times 4$  matrix and satisfies the following equation (see Fig.~\ref{fig:ladder}):
\ber \label{eq:vertex}
\Lambda(\qv,i \nu_{m}) &=&  \Gamma^{2} - V~\tau^3\otimes \openone_{\sigma} \Bigg\{\frac{1}{\beta}\sum_{\bm k,n}G_0({\bm k}^+, i\epsilon^+_{n}) \nonumber \\
&\times& \Lambda(\qv,  i\nu_{m}) G_0({\bm k}^-, i\epsilon^-_{n})\Bigg\}~\tau^3 \otimes \openone_{\sigma}~. 
\eer
In the $q \to 0$ limit, after some lengthy but straightforward algebra, Eq.~(\ref{eq:vertex}) yields
\be
\Lambda(0,i \nu_{m}) = \frac{\openone_\tau \otimes \sigma^2}{\displaystyle 1- \frac{V}{2}\chi_{\sigma^2 \sigma^2}(0,i \nu_{m})}~.
\ee
In the ladder approximation, the interacting in-plane spin-susceptibility in the long-wavelength limit is thus given by
\be \label{eq:chiy-final}
{\widetilde \chi}_{\sigma^2 \sigma^2} (0, i \nu_{m}) = \frac{\chi_{\sigma^2 \sigma^2} (0,i \nu_{m})}{\displaystyle 1- \frac{V}{2}\chi_{\sigma^2 \sigma^2}(0,i \nu_{m})}~.
\ee
Since collective modes are isolated poles in the dynamical response function ${\widetilde \chi}_{\sigma^2 \sigma^2} (0, i \nu_{m} \to \omega + i 0^+)$ (located infinitesimally below the real-frequency axis), 
Eq.~(\ref{eq:chiy-final}) reproduces the condition given above in Eq.~(\ref{polesexplicit}).

\section{Normal Rashba Gas}
\label{Appendix: normal}

In this Section we report explicit expressions for the real and imaginary parts of the in-plane dynamical spin susceptibility $\left.\chi_{\sigma^2\sigma^2}(0,\omega)\right|_{\Delta =0}$ of a normal (non-superconducting) Rashba 2DEG in the absence of electron-electron interactions [see panel a) of Fig.~1 in the main text].

At zero temperature and in the absence of superconductivity we find:
\ber \label{eq: chi00}
\left.\chi_{\sigma^2\sigma^2}(0,\omega)\right|_{\Delta = 0} &=& \frac{1}{4 \pi} \int_{k_{\rm F}, +}^{k_{\rm F},-} k dk
\Big[ \frac{1}{\omega + 2 \alpha k + i 0^+} \nonumber \\
&-& \frac{1}{\omega - 2 \alpha k + i 0^+}\Big] ~,
\eer
where $k_{{\rm F}, \pm} = (2 m \mu + m^{2} \alpha^{2})^{1/2} \mp m \alpha$ is the Fermi wave-vector of the minority (majority) Rashba band.  Performing the integration in Eq.~(\ref{eq: chi00}), we find that 
the real and imaginary parts of $\left.\chi_{\sigma^2\sigma^2}(0,\omega)\right|_{\Delta = 0}$ are given by 
\ber\label{eq: Rechi1}
\Re e~\left.\chi_{\sigma^2\sigma^2}(0,\omega)\right|_{\Delta = 0}& = &\frac{m}{2 \pi} \Bigg(1+ \frac{\omega}{8 m \alpha^{2}}  \nonumber \\
& \times& \ln\left| \frac{\omega+ 2 \alpha k_{\rm F, +}}{\omega- 2 \alpha k_{\rm F, +}}~\frac{\omega- 2 \alpha k_{\rm F, -}}{\omega+ 2 \alpha k_{\rm F, -}}  \right| \Bigg)  \nonumber \\
\eer
and
\ber\label{eq: Imchi1}
\Im m~\left.\chi_{\sigma^2\sigma^2}(0,\omega)\right|_{\Delta = 0} &=& \frac{\omega}{16 \alpha^{2}} \Theta(\omega - 2 \alpha k_{\rm F,+}) \nonumber \\
& \times& \Theta (2 \alpha k_{\rm F,-} - \omega)~.
\eer
Eqs.~(\ref{eq: Rechi1}) and (\ref{eq: Imchi1}) agree with Eqs.~($7$) and~($10$) in Ref.~\onlinecite{Lopez} (after setting to zero the Dresselhaus spin-orbit coupling constant in their results).
\end{appendix} 


\begin{thebibliography}{77}
\bibitem{qfactor}
	The quality factor of the oscillations -- limited by radiative damping --  is estimated to be $Q=c/(\omega r_0)$ where $r_0= e^2/(mc^2) \sim 3\times 10^{-15}~{\rm m}$ 
	is the classical electron radius, $\omega$ is the frequency, and $c$ is the speed of light.  This gives a value of $Q$ in excess of $10^{10}$.
\end{thebibliography}
\end{document}